\documentclass[draftclsnofoot,onecolumn]{IEEEtran}
\usepackage{pifont}
\usepackage{graphicx}
\usepackage{fancyhdr}
\usepackage{amsmath,amssymb,amstext,amsfonts}
\usepackage{slashbox}
\usepackage{bm}
\usepackage{algorithm}
\usepackage{array}
\usepackage{cite}

\usepackage{subfigure}
\usepackage{balance}
\usepackage{algorithmicx}
\usepackage{algpseudocode}

\usepackage{braket}
\usepackage{tabulary}
\usepackage{diagbox}
\usepackage{amsthm}
\usepackage{color}
\usepackage{placeins}

\usepackage{tikz}
\usetikzlibrary{shapes,arrows,positioning,fit,backgrounds,shadows}

\makeatletter
\renewcommand{\maketag@@@}[1]{\hbox{\m@th\normalsize\normalfont#1}}%
\makeatother

\begin{document}

\title{SCI-D$^2$NN: An Optimization Framework for OAM-Multiplexed FSO Communications}

\author{Rui~Deng, Renzhi~Yuan$^{*}$, \IEEEmembership{Member,~IEEE}, Xinyi Chu, Siming Wang, Chengzhi Liu, 

Zehao He, Haifeng Yao, and Mugen~Peng, \IEEEmembership{Fellow,~IEEE}.
\thanks{
Rui~Deng, Renzhi~Yuan, Xinyi Chu, Siming Wang, and Mugen~Peng are with the State Key Laboratory of Networking and Switching Technology, Beijing University of Posts and Telecommunications, Beijing 100876, China, and also with Beijing Key Laboratory of Convergent Communications and Networking Technologies in LEO Satellite Systems; Chengzhi Liu is with the School of Artificial Intelligence and Computer Science, North China University of Technology, Beijing 100144, China; Zehao He is with Department of Physics, Capital Normal University, Beijing 100048, China; Haifeng Yao is with the School of Optics and Photonics, Beijing Institute of Technology, Beijing 100081, China,  and also with angtze Delta Region Academy of Beijing Institute of Technology, Jiaxing 314019, China. 

Corresponding author: Renzhi~Yuan.

This work is supported by Beijing Natural Science Foundation under Grant No. 4262010, and also by the National Key Laboratory on Near-Surface Detection under Grant No. NKLNSD02202501.

This manuscript is currently under peer review.
}}

\maketitle

\begin{abstract}
Orbital angular momentum (OAM) multiplexing can increase the capacity of free-space optical (FSO) communications, but its detection performance is strongly affected by impairments such as atmospheric turbulence, transmitter pointing errors, and photodetection noise. The diffractive deep neural network (D$^2$NN) can be used as an all-optical front end to mitigate turbulence-induced distortions before detection. However, existing D$^2$NN compensation schemes are not specifically optimized for communication detection. In this paper, we propose a supervised contrastive inspired D$^2$NN (SCI-D$^2$NN) framework for improving the detection performance of OAM-multiplexed FSO communications under these impairments. The proposed framework introduces two training branches: a projection branch that maps the optical field to low-dimensional decision domain samples, and a label branch that provides supervised labels to impose a separation constraint among decision domain samples. {In addition, we characterize complex-amplitude crosstalk to obtain the receiver observation vector and formulate two detection schemes, namely single-port profile-likelihood detection and joint maximum-likelihood (ML) detection.} We further design two SCI-D$^2$NN training losses called Bhattacharyya distance (BD) based loss and the ML based loss to improve decision domain separability and mitigate detection-performance degradation. Numerical results show that SCI-D$^2$NN achieves more than a 3-dB improvement in bit error rate (BER) over the conventional D$^2$NN baseline in most transmit-power regions. The BD based loss gives the lowest BER under different system parameters and provides more than a 10-dB BER improvement over the baseline in the high transmit power region.
\end{abstract}


\section{Introduction}
\subsection{Background and Motivation}
Free-space optical (FSO) communications use optical waves as the information-bearing medium and have attracted extensive attention owing to higher bandwidth, low-latency, and flexible deployment. With the development of high-capacity applications such as the Internet of Things, artificial intelligence, and satellite laser communications, modern communication systems are required to support higher data rates and larger transmission capacities~\cite{Alimi2024RevolutionizingFSO}. In conventional optical communication systems, dimensions such as frequency, wavelength, and polarization have been widely exploited~\cite{Wang2018Metasurfaces}. These constraints motivate the exploration of additional physical dimensions for multiplexing. Optical beams carrying orbital angular momentum (OAM) have been widely studied as such a spatial modal resource in FSO communications~\cite{Allen1992OAM,Wang2022OAMBeyondFSO}. An OAM beam exhibits a helical phase front, and modes with different topological charges are orthogonal under ideal propagation conditions. Therefore, OAM-based FSO communication is a promising approach for expanding the spatial multiplexing dimension and improving the transmission capability of optical wireless links. However, the performance of OAM-multiplexed FSO communication is degraded by impairments such as atmospheric turbulence, transmitter pointing errors, and photodetection noise. Existing compensation schemes either rely on electrical-domain processing, which may introduce additional communication latency, or use fast compensation in the optical domain without explicitly targeting the communication detection process. It is desirable to develop an all-optical compensation scheme designed for communication-oriented detection performance improvement.

\subsection{Related Works}
OAM-based FSO communications have been investigated in many studies. For example, OAM multiplexing has been used to transmit parallel data channels over a coaxial FSO link~\cite{Wang2012OAMFSO}. OAM modes have also been combined with four-level pulse-amplitude modulation (PAM-4), OAM shift keying, and OAM index modulation to increase the information carried by modal states~\cite{Liu2016PAM4OAM,Kai2017OAMSK,Basar2018OAMIM,Zhang2019OAMIndexMapping}. These studies show that OAM modes provide an effective spatial degree of freedom for enhancing the transmission capability of FSO links. However, the above works do not explicitly consider atmospheric turbulence, which breaks modal orthogonality and degrades the performance of OAM-FSO communication.

To mitigate turbulence-induced performance degradation, various electrical-domain compensation methods have been investigated, including adaptive optics based compensation~\cite{Ren2014AOCompensation,Chang2021LowComplexityAO,Chang2022AOCompensation,Chang2024NonprobeAO,Su2023AOLG}, iterative wavefront correction~\cite{Xie2015SPGDOAM,Fu2016GSTurbulence}, and deep learning assisted distortion compensation~\cite{Liu2019DeepLearningCompensation,Lu2020JointCNN}. However, these methods usually require iterative computation, optoelectronic conversion, or digital signal processing, which increase computational cost and introduce additional communication latency in high-speed FSO links.

All-optical compensation based on the diffractive deep neural network (D$^2$NN) introduced by Lin \emph{et al.}~\cite{Lin2018AllOpticalD2NN} provides a possible route to reducing communication latency.  D$^2$NN has been widely used in optical computing and processing tasks, including image classification and object recognition~\cite{Rahman2021EnsembleD2NN,Mengu2021InvariantD2NN,Bai2023SinglePixelD2NN}, computational imaging and optical computing~\cite{Luo2019TaskSpecificD2NN,Chen2021VisibleD2NN,Mengu2022OverlappingPhase}, phase retrieval~\cite{Goi2022ZernikeD2NN}, and optical field processing~\cite{Jia2022D2NNDelay,Zhan2022DDNNAO}. Recently, D$^2$NN-based methods have also been investigated for turbulence-distorted OAM beams. Zhan et al. proposed a D$^2$NN-based adaptive optics (AO) scheme that predicts turbulence phase screens from the distorted intensity distribution of OAM beams to achieve conjugate phase compensation~\cite{Zhan2022DDNNAO}. The same line of work further introduced a photoelectric hybrid deep neural network (PHDNN), where a convolutional neural network (CNN) reconstructs phase screens through Zernike-coefficient prediction~\cite{Zhan2026PHDNNAO}. {These schemes improve phase compensation, but still rely on electronic-domain inference and subsequent phase-screen loading.} In contrast, Jia et al. proposed an end-to-end D$^2$NN architecture for near-zero time-delay restoration of distorted optical fields in the optical domain~\cite{Jia2022D2NNDelay}. However, the training objectives of these D$^2$NN architectures are not designed for communication performance optimization.

In practical OAM-multiplexed FSO systems, atmospheric turbulence distorts the wavefront, while transmitter pointing errors introduce receiver-plane misalignment~\cite{Wang2021MisalignedOAM,Gong2021PointingOAMSK}; both effects reduce modal orthogonality and increase inter-mode crosstalk. Moreover, the receiver noise after photodetection directly affects the decision statistics used for symbol detection~\cite{Kahn1997WirelessIR,Khalighi2014FSOSurvey}. Therefore, a low-latency D$^2$NN framework should be designed for the communication detection task such that the receiver-side detection performance can be improved under turbulence-induced distortion, pointing-error-induced misalignment, and photodetection noise.

\subsection{Contributions}
{In this paper, we propose a supervised contrastive inspired D$^2$NN (SCI-D$^2$NN) framework for improving detection performance in OAM-multiplexed FSO communication under atmospheric turbulence, transmitter pointing errors, and photodetection noise.} We first establish an OAM-FSO channel model that accounts for turbulence and pointing errors. Then, we characterize the complex-amplitude crosstalk to obtain the receiver observation vector. Based on this vector, the detection schemes and corresponding detection criteria are formulated. Moreover, we propose a SCI-D$^2$NN framework with two training branches and design two loss functions to improve the separability of decision domain samples. Numerical simulations are carried out to assess the detection performance of the proposed framework. The main contributions are summarized as follows:

\begin{itemize}
    \item We propose the SCI-D$^2$NN training framework by incorporating a projection branch and a label branch into the training phase to improve detection performance. Specifically, the projection branch maps the optical field to low-dimensional decision-domain samples, thereby simplifying the training task of SCI-D$^2$NN. The label branch uses the transmitted symbol states as supervised labels to impose separation constraints among decision-domain samples, enabling the SCI-D$^2$NN to improve their decision-domain separability.

    \item {We characterize the crosstalk that is evaluated based on the receiver-side complex field for OAM-multiplexed communication to obtain the receiver observation vector.} Based on this vector, we formulate single-port profile-likelihood (PL) detection and joint maximum-likelihood (ML) detection, and derive the corresponding bit error rate (BER) and symbol error rate (SER) as evaluation criteria.

    \item To improve the communication detection performance of OAM-multiplexed FSO communications under atmospheric turbulence, transmitter pointing errors, and photodetection noise, we design two SCI-D$^2$NN training losses: the Bhattacharyya distance (BD) based loss and the ML based loss.

    \item We demonstrated that the proposed SCI-D$^2$NN framework achieves more than a 3-dB BER improvement over the conventional D$^2$NN baseline in most transmit-power regions. SCI-D$^2$NN enables this improvement by reducing the dimensionality of the training samples and aligning the optimization objective with the final detection performance.

    \item We demonstrated that the BD based loss achieves the lowest BER under different system parameters, i.e., symbol rate, turbulence strength, diffractive layer, and pointing error. By most effectively improving the separability of samples in the weak-separation tail, it provides more than a 10-dB BER improvement over the conventional D$^2$NN baseline in the high transmit power region.
\end{itemize}

The remainder of this paper is organized as follows. Section~\ref{sec:system_channel_model} presents the system and receiver models. {Section III formulates the single-port PL detection and joint ML detection schemes based on the receiver observation vector.} Section~\ref{sec:sci_d2nn_training} introduces the proposed SCI-D$^2$NN training framework and the baseline model. Section~\ref{sec:numerical_results} provides the numerical results and performance analysis. Section~\ref{sec:conclusion} concludes this paper.

\begin{figure*}[t]
    \centering
    \includegraphics[width=0.75\textwidth]{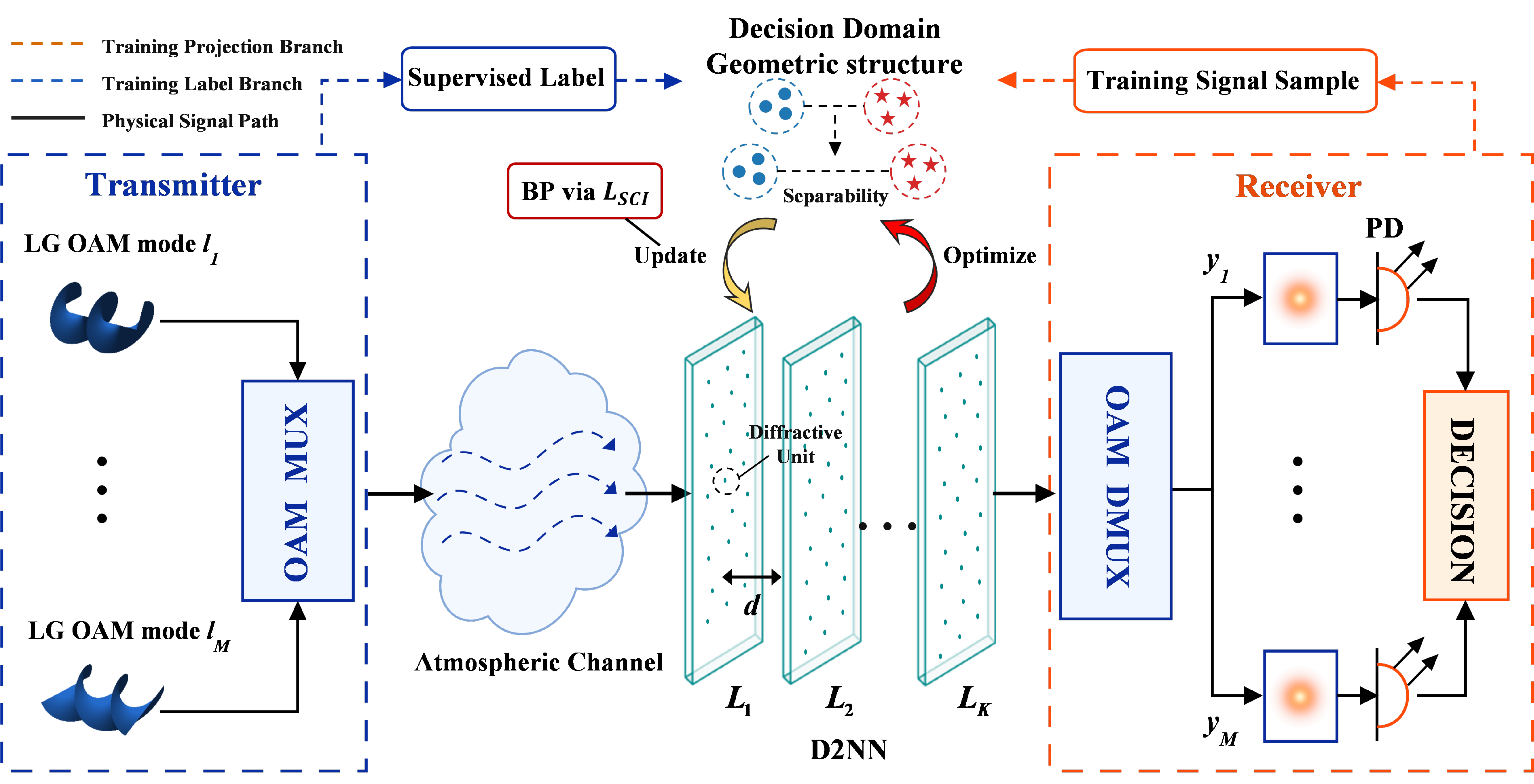}
    \vspace{-0.6em}
    \caption{OAM-multiplexed FSO communication system based on SCI-D$^2$NN. (BP: backpropagation; PD: photodetection; MUX: multiplexer; DMUX: demultiplexer)}
    \label{fig:system_framework}
\end{figure*}

\section{OAM-Multiplexed FSO Communication System Model in Turbulent Channels}
\label{sec:system_channel_model}

The OAM-multiplexed FSO communication system and the training structure of the proposed SCI-D$^2$NN framework are shown in Fig.~\ref{fig:system_framework}. At the transmitter of physical signal path, multiple Laguerre-Gaussian (LG) OAM branches are coherently combined to produce a coaxially multiplexed optical field. The multiplexed field propagates through the atmospheric channel and is affected by turbulence and pointing errors before entering the SCI-D$^2$NN optical front end. At the receiver, the SCI-D$^2$NN output field is demultiplexed by modal projection, and the projected components are photodetected to obtain the receiver observation vector used for the final decision. During training, the SCI-D$^2$NN framework introduces two auxiliary branches. The training label branch provides supervised labels, while the training projection branch provides receiver-side training signal samples. Together, these two branches construct the decision-domain geometric structure used by the SCI-D$^2$NN loss. Leveraging this geometry, the SCI-D$^2$NN loss function is formulated to improve the separability of samples in the decision domain, and the SCI-D$^2$NN parameters are updated through backpropagation.

\subsection{OAM-Multiplexed Signals}

An optical field carrying an OAM of $\ell\hbar$ per photon has a helical phase front, whose azimuthal phase term can be expressed as $\exp(j\ell\phi)$. Here, $j=\sqrt{-1}$, $\hbar$ is the reduced Planck constant, $\ell$ is the topological charge, and $\phi$ is the azimuth angle~\cite{Allen1992OAM}. Under ideal propagation, OAM modes associated with different topological charges are mutually orthogonal. In the cylindrical coordinate system $(\rho,\phi,z)$, the spatial distribution of an LG beam with radial index $p$ and topological charge $\ell$ is written as~\cite{Doster2016LGBG}
\begin{align}
u_{p,\ell}(\rho,\phi,z)
&=\sqrt{\frac{2p!}{\pi(p+|\ell|)!}}\frac{1}{w(z)}
\left(\frac{\sqrt{2}\rho}{w(z)}\right)^{|\ell|}
L_p^{|\ell|}\left(\frac{2\rho^2}{w^2(z)}\right) \notag\\
&\quad \times \exp\left(-\frac{\rho^2}{w^2(z)}\right)
\exp\left(-\frac{jk\rho^2z}{2(z^2+z_R^2)}\right) \notag\\
&\quad \times
\exp\left[j(2p+|\ell|+1)\arctan\left(\frac{z}{z_R}\right)\right] \notag\\
&\quad \times \exp(-j\ell\phi),
\label{eq:lg_mode}
\end{align}
where $\rho$ is the radial distance, $w(z)=w_0\sqrt{1+(z/z_R)^2}$ is the beam radius at distance $z$, $w_0$ is the beam waist, $z_R=\pi w_0^2/\lambda$ is the Rayleigh range, $\lambda$ is the operating wavelength, $k=2\pi/\lambda$ is the propagation constant, and $L_p^{|\ell|}(\cdot)$ is the generalized Laguerre polynomial.  In this paper, only the radial index $p=0$ is considered, and we write $u_{\ell}(\rho,\phi,z)\triangleq u_{0,\ell}(\rho,\phi,z)$. The finite OAM mode set used for multiplexing is denoted by $S=\{\ell_1,\ell_2,\ldots,\ell_M\}$, where $M$ is the number of multiplexed OAM channels. The orthogonality of normalized OAM fields over the ideal transverse plane $\Omega_{\perp}$ is~\cite{Huang2018SpatialMode}
\begin{equation}
\int_{\Omega_{\perp}}u_{\ell_m}(\rho,\phi,z)u_{\ell_n}^{*}(\rho,\phi,z)\,\mathrm{d} A
=\delta_{mn},
\label{eq:oam_orthogonality}
\end{equation}
where $\mathrm{d} A=\rho\,\mathrm{d}\rho\,\mathrm{d}\phi$ and $\delta_{mn}$ is the Kronecker delta function. The orthogonality relation in \eqref{eq:oam_orthogonality} provides the  modal-projection basis for the receiver. During propagation through the FSO channel, atmospheric turbulence and pointing errors disturb the wavefront and receiver-plane alignment of OAM beams, causing optical power to leak from the intended OAM mode branches into other branches. The resulting modal leakage gives rise to crosstalk among OAM channels that would be separated under ideal propagation.

In an OAM-multiplexed FSO communication system, all OAM branches are modulated from the same coherent optical source and are coherently superposed at the transmit plane. In one symbol interval, the modulation symbols loaded onto the OAM modes are collected as
\begin{equation}
\mathbf{c}=[c_1,c_2,\ldots,c_M]^T.
\label{eq:mod_symbol}
\end{equation}
Here, $c_m$ is the normalized modulation symbol carried by OAM mode $\ell_m$. Let $P_m$ denote the transmit power of the $m$th branch. The OAM-multiplexed transmit field is
\begin{equation}
E_{\mathrm{tx}}(\rho,\phi;\mathbf{c})
=\sum_{m=1}^{M}\sqrt{P_m}\,c_m u_{\ell_m}(\rho,\phi,z_0).
\label{eq:tx_field_general}
\end{equation}
The average transmitted power over one symbol interval is defined as
\begin{equation}
P_{\mathrm{avg}}
=
\mathbb{E}_{\mathbf{c}}
\left[
\sum_{m=1}^{M}P_m |c_m|^2
\right],
\label{eq:average_transmitted_power}
\end{equation}
where the expectation is taken over the transmitted symbol vector $\mathbf{c}$. The transmit field in \eqref{eq:tx_field_general} is then used as the channel input for the atmospheric propagation model.

\subsection{Turbulent Effects and Pointing Errors for OAM-Multiplexed Signals}

In the considered FSO communication link, the propagation from the transmit field to the SCI-D$^2$NN input field in the $s$th channel realization is denoted by
\begin{equation}
E_{\mathrm{turb}}^s=H_s(E_{\mathrm{tx}}),
\label{eq:channel_mapping}
\end{equation}
where $H_s(\cdot)$ is the propagation operator. In the numerical implementation, it includes turbulence propagation modeled by multiple phase screens and the receiver-plane displacement caused by pointing errors.

Let $d$ be the propagation distance, $N_p$ be the number of phase screens, and $\Delta z=d/N_p$ be the spacing between adjacent screens. Let $E_{\nu}^{s}(\rho,\phi;\mathbf{c})$ denote the complex-field sample after the $\nu$th propagation step in the $s$th turbulence realization, where $\nu=0,1,\ldots,N_p$, with initial condition $E_{0}^{s}(\rho,\phi;\mathbf{c})=E_{\mathrm{tx}}(\rho,\phi;\mathbf{c})$. The multi-phase-screen recursion is
\begin{equation}
E_{\nu}^{s}
=G_{\Delta z}\left[E_{\nu-1}^{s}\exp\left(j\varphi_{\nu}^{s}\right)\right],
\quad \nu=1,\ldots,N_p,
\label{eq:phase_screen}
\end{equation}
where $\varphi_{\nu}^{s}(\rho,\phi)$ is the random phase perturbation introduced by the $\nu$th phase screen, and $G_{\Delta z}[\cdot]$ is the free-space propagation operator over distance $\Delta z$. This operator is implemented using the angular spectrum method~\cite{Matsushima2009BandLimitedASM}.

The random phase perturbation $\varphi_{\nu}^{s}(\rho,\phi)$ in \eqref{eq:phase_screen} is generated from the turbulence statistics of the corresponding thin phase screen. We use the von Karman refractive-index power spectrum~\cite{Karman1948Turbulence} to specify these statistics. For spatial-frequency magnitude $\kappa$, the refractive-index fluctuation spectrum is
\begin{equation}
\Phi_n(\kappa)=0.033C_n^2
\frac{\exp(-\kappa^2/\kappa_m^2)}
{(\kappa^2+\kappa_0^2)^{11/6}},
\label{eq:von_karman}
\end{equation}
where $C_n^2$ is the refractive-index structure constant, $\kappa_0=2\pi/L_0$, $\kappa_m=5.92/l_0$, and $L_0$ and $l_0$ are the outer and inner scales of turbulence. Accordingly, \eqref{eq:von_karman} determines the spectrum used to synthesize $\varphi_{\nu}^{s}(\rho,\phi)$, while \eqref{eq:phase_screen} describes how the sampled phase perturbation is applied to the optical field during split-step propagation.

In addition to turbulence-induced wavefront distortion, pointing errors introduce random angular misalignment between the transmitted beam axis and the receiver optical axis. Let the transverse angular error vector be
\begin{equation}
\boldsymbol{\theta}_{\perp}
=
[\theta_x,\theta_y]^T,
\label{eq:transverse_angular_error}
\end{equation}
where $\theta_x$ and $\theta_y$ are the angular errors along two orthogonal transverse directions. Its polar parametrization is $\theta=\|\boldsymbol{\theta}_{\perp}\|$ and $\psi=\operatorname{atan2}(\theta_y,\theta_x)$, or equivalently $\theta_x=\theta\cos\psi$ and $\theta_y=\theta\sin\psi$. If $\theta_x$ and $\theta_y$ follow Gaussian distributions with means $\mu_x$ and $\mu_y$ and variances $\sigma_x^2$ and $\sigma_y^2$, respectively, the probability density function of $\theta$ can be written as~\cite{BoludaRuiz2016Beckmann}
\begin{align}
f_{\theta}(\theta)
&=
\frac{\theta}{2\pi\sigma_x\sigma_y}
\int_{0}^{2\pi}
\exp\Bigg[
-\frac{(\theta\cos\psi-\mu_x)^2}{2\sigma_x^2}
\notag\\
&\qquad
-\frac{(\theta\sin\psi-\mu_y)^2}{2\sigma_y^2}
\Bigg]\mathrm{d}\psi,
\quad \theta\geq 0.
\label{eq:pointing_pdf_general}
\end{align}
When $\mu_x=\mu_y=0$ and $\sigma_x=\sigma_y=\sigma_{\theta}$, the pointing-error angle follows the Rayleigh distribution,
\begin{equation}
f_{\theta}(\theta)
=
\frac{\theta}{\sigma_{\theta}^{2}}
\exp\left(-\frac{\theta^{2}}{2\sigma_{\theta}^{2}}\right),
\quad \theta\geq 0.
\label{eq:pointing_pdf_rayleigh}
\end{equation}
The numerical study uses this zero-mean isotropic pointing-error model unless otherwise specified.
For the $s$th channel realization, the sampled transverse angular error produces a receiver-plane displacement
\begin{equation}
\Delta_x^s=d\theta_{x,s}=d\theta_s\cos\psi_s,\qquad
\Delta_y^s=d\theta_{y,s}=d\theta_s\sin\psi_s,
\label{eq:pointing_displacement}
\end{equation}
under the small-angle approximation. Therefore, the distorted complex field entering the SCI-D$^2$NN after turbulence propagation and pointing-error displacement is modeled as
\begin{equation}
E_{\mathrm{turb}}^s(x,y;\mathbf{c})
=E_{N_p}^{s}(x-\Delta_x^s,y-\Delta_y^s;\mathbf{c}).
\label{eq:turb_field}
\end{equation}
Equations \eqref{eq:channel_mapping}, \eqref{eq:phase_screen}, \eqref{eq:pointing_pdf_rayleigh}, \eqref{eq:pointing_displacement}, and \eqref{eq:turb_field} define the distorted complex field used as the SCI-D$^2$NN input.

After optical front end processing, the SCI-D$^2$NN maps the distorted input field to the SCI-D$^2$NN output field at the receiver plane. This input-output relation is written in compact form as
\begin{equation}
E_{\mathrm{net}}=F_{\Theta}(E_{\mathrm{turb}}),
\label{eq:d2nn_mapping}
\end{equation}
where $F_{\Theta}(\cdot)$ is the optical mapping operator induced by multi-layer diffractive modulation and free-space propagation, and $\Theta$ denotes the trainable amplitude and phase parameters.

\subsection{Mode Crosstalk of OAM Modes}

Accurate crosstalk characterization is important for detection analysis in OAM-multiplexed FSO communication. Existing OAM crosstalk analyses often describe modal leakage in the power domain and do not account for the phase relations among leaked components~\cite{Amhoud2020UnifiedFading}. Since OAM-multiplexed signals are coherently superposed before photodetection, we characterize crosstalk in the complex-amplitude domain.

{To construct the receiver-side crosstalk matrix, for the $s$th channel realization and fixed SCI-D$^2$NN parameters $\Theta$, let $E_{\mathrm{net},m}^{(s)}$ denote the output field of the cascade when the $m$th OAM branch is launched alone with unit amplitude.} Here, $E_{\mathrm{turb},m}^{(s)}$ is the corresponding turbulence-distorted input field entering the SCI-D$^2$NN. Projecting $E_{\mathrm{net},m}^{(s)}$ onto the normalized receiver OAM basis $u_{\ell_r}$ gives the complex-amplitude crosstalk coefficient
\begin{equation}
A^{(s)}_{r,m}
=
\left\langle u_{\ell_r},E_{\mathrm{net},m}^{(s)}\right\rangle
=
\int u_{\ell_r}^{*}(\rho,\phi,z_{\mathrm{rx}})
E_{\mathrm{net},m}^{(s)}(\rho,\phi,z_{\mathrm{rx}})\,\mathrm{d} A,
\label{eq:complex_crosstalk}
\end{equation}
where $\mathrm{d} A$ denotes the differential area element on the receiver transverse plane.
{
Thus, $A_{r,m}^{(s)}$ is the complex amplitude obtained at receiver port $r$ after modal projection. Collecting these coefficients gives the receiver-side complex-amplitude crosstalk matrix for the $s$th channel realization,
\begin{equation}
\mathbf{A}_s=
\begin{bmatrix}
A^{(s)}_{1,1} & A^{(s)}_{1,2} & \cdots & A^{(s)}_{1,M}\\
A^{(s)}_{2,1} & A^{(s)}_{2,2} & \cdots & A^{(s)}_{2,M}\\
\vdots & \vdots & \ddots & \vdots\\
A^{(s)}_{M,1} & A^{(s)}_{M,2} & \cdots & A^{(s)}_{M,M}
\end{bmatrix}.
\label{eq:coupling_matrix}
\end{equation}
The $m$th column of $\mathbf{A}_s$ contains the complex-amplitude responses at all receiver ports produced by the $m$th transmitted OAM branch.
}
The diagonal entries of $\mathbf{A}_s$ describe the self-mode coupling responses, whereas the off-diagonal entries describe complex-amplitude leakage from transmitted OAM branches to other receiver ports. Unlike power-domain crosstalk metrics that retain only $|A^{(s)}_{r,m}|^2$, the complex-amplitude matrix $\mathbf{A}_s$ preserves the relative phase among leakage components.

{
For a transmitted modulation state $\mathbf{c}$, the complex-amplitude components from different OAM branches are coherently superposed at each receiver port. Because channel propagation, SCI-D$^2$NN processing, and modal projection are linear in the launched field for fixed $s$, the projected complex-amplitude vector is
\begin{equation}
\mathbf{g}_s(\mathbf{c})
=
\mathbf{A}_s\mathbf{P}^{1/2}\mathbf{c},
\label{eq:complex_amplitude_state}
\end{equation}
where $\mathbf{P}^{1/2}
=\operatorname{diag}(\sqrt{P_1},\sqrt{P_2},\ldots,\sqrt{P_M})$ contains the branch transmit-power factors. Thus, the matrix multiplication in \eqref{eq:complex_amplitude_state} performs the coherent superposition of the projected complex-amplitude responses from all transmitted branches.
}
The corresponding receiver-side intensity vector is
\begin{equation}
\mathbf{I}_s(\mathbf{c})
=
|\mathbf{g}_s(\mathbf{c})|^2
=
|\mathbf{A}_s\mathbf{P}^{1/2}\mathbf{c}|^2,
\label{eq:intensity_state}
\end{equation}
where the absolute value and square are applied entry by entry. For receiver port $r$, the corresponding intensity is
\begin{equation}
I_{s,r}(\mathbf{c})
=
\left|
\sum_{m=1}^{M}
A^{(s)}_{r,m}\sqrt{P_m}c_m
\right|^2.
\label{eq:port_intensity}
\end{equation}
Expanding the square gives
\begin{align}
I_{s,r}(\mathbf{c})
&=
\sum_{m=1}^{M}
P_m |A^{(s)}_{r,m}|^2 |c_m|^2
\notag\\
&\quad
+2\sum_{m<n}
\sqrt{P_mP_n}\,
\operatorname{Re}\!\left\{
A^{(s)}_{r,m}
\left(A^{(s)}_{r,n}\right)^{*}
c_m c_n^{*}
\right\}.
\label{eq:intensity_expansion}
\end{align}
The first term in \eqref{eq:intensity_expansion} is determined by the modal leakage power associated with individual OAM branches, whereas the second term is produced by the coherent cross terms among different branches. Hence, the receiver-side intensity vector depends on both the leakage magnitudes and the relative phases captured by $\mathbf{A}_s$.

This section characterizes complex-amplitude crosstalk and derives the corresponding receiver-side intensity vector used in the subsequent detection analysis.

\section{Single-Port PL and Joint ML Detection}

This paper focuses on using the SCI-D$^2$NN framework to mitigate the degradation of detection performance caused by atmospheric turbulence and pointing errors. For simplicity, on-off keying (OOK) is used as the modulation format for the following analysis.

\subsection{Observation Vector for Detection}
Under OOK modulation, the modulation symbol on branch $m$ is set as $c_m=b_m$. The joint bit state in one symbol interval is
\begin{equation}
\mathbf{b}=[b_1,b_2,\ldots,b_M]^T,\qquad b_m\in\{0,1\},
\label{eq:ook_state}
\end{equation}
where $b_m$ denotes the on-off state of the $m$th OAM branch. The candidate joint-state set is $\mathcal{B}=\{0,1\}^{M}$ with $|\mathcal{B}|=2^M$.

After photodetection, the receiver-side intensity vector $\mathbf{I}(\mathbf{b})$ is converted into the receiver observation vector. For a given joint bit state $\mathbf{b}$, the observation model is
\begin{equation}
\mathbf{Y}=R\mathbf{I}(\mathbf{b})+\boldsymbol{\eta},
\label{eq:observation}
\end{equation}
where $R$ is the photodetector responsivity and $\boldsymbol{\eta}$ is the receiver noise vector. Due to modal crosstalk, the power at one receiver port is contributed by its corresponding OAM branch and leakage from other branches. As a result, the received intensity at each port varies with the joint transmitted state, leading to state-dependent shot-noise variance. We consider independent thermal noise and shot noise at different receiver ports. When the number of photoelectrons collected within the receiver bandwidth is sufficiently large, the Poisson shot noise can be approximated by a Gaussian distribution~\cite{ramaswami2009optical}. The noise variance at port $r$ is given by
\begin{equation}
\sigma_r^2(I_r)=
\frac{4k_BT B_e}{R_L}+2q_eB_eRI_r,
\label{eq:noise_variance}
\end{equation}
where $k_B$ is the Boltzmann constant, $T$ is the absolute temperature, $B_e$ is the equivalent electrical bandwidth, $R_L$ is the load resistance, and $q_e$ is the electron charge. Since $\sigma_r^2(I_r)$ depends on the received intensity, different joint states lead to different noise variances. Thus, conditioned on $\mathbf{b}$, the observation vector follows
\begin{equation}
\mathbf{Y}\mid \mathbf{b}
\sim \mathcal{N}\left(R\mathbf{I}(\mathbf{b}),\boldsymbol{\Sigma}(\mathbf{b})\right),
\label{eq:hetero_observation}
\end{equation}
where the state-dependent covariance matrix is
\begin{equation}
\boldsymbol{\Sigma}(\mathbf{b})
=\operatorname{diag}\left(
\sigma_1^2(I_1(\mathbf{b})),\sigma_2^2(I_2(\mathbf{b})),\ldots,
\sigma_M^2(I_M(\mathbf{b}))
\right).
\label{eq:noise_covariance}
\end{equation}

Based on the observation vector $\mathbf{Y}$, we consider single-port PL detection using the scalar observation of one selected receiver port and joint ML detection using the full multi-port observation vector.

\subsection{Single-Port PL Detection}

Consider receiver port $r$. The single-port detector estimates only the corresponding OAM branch bit $b_r$ from the scalar observation $Y_r$ and treats the remaining OAM branch bits as unknown nuisance states. The candidate joint states are divided according to the value of the bit on branch $r$:
\begin{equation}
\mathcal{B}_{r,1}=\{\mathbf{b}\in\mathcal{B}:b_r=1\},\qquad
\mathcal{B}_{r,0}=\{\mathbf{b}\in\mathcal{B}:b_r=0\}.
\label{eq:single_state_sets}
\end{equation}

For port $r$ and hypothesis $\beta\in\{0,1\}$, define the corresponding candidate intensity set as
\begin{equation}
\mathcal{I}_{s,r,\beta}
=
\{I_{s,r}(\mathbf{a})\mid \mathbf{a}\in\mathcal{B}_{r,\beta}\}.
\label{eq:single_candidate_intensity_set}
\end{equation}
For the $s$th channel realization, $I_{s,r}(\mathbf{a})$ denotes the noiseless receiver-side intensity at port $r$ when the candidate joint state is $\mathbf{a}\in\mathcal{B}$. The corresponding scalar observation model is
\begin{equation}
Y_r\mid \mathbf{a},s
\sim
\mathcal{N}\left(
RI_{s,r}(\mathbf{a}),
\sigma_r^2(I_{s,r}(\mathbf{a}))
\right).
\label{eq:single_conditional_observation}
\end{equation}
For a scalar observation $y_r$, the log-likelihood score of candidate state $\mathbf{a}$ at port $r$ is, up to an additive constant independent of $\mathbf{a}$,
\begin{equation}
\mathcal{L}_{s,r}(\mathbf{a};y_r)
=
-\frac{1}{2}
\left[
\frac{
\left(y_r-RI_{s,r}(\mathbf{a})\right)^2
}{
\sigma_r^2(I_{s,r}(\mathbf{a}))
}
+
\log\sigma_r^2(I_{s,r}(\mathbf{a}))
\right].
\label{eq:single_candidate_log_likelihood}
\end{equation}

Since the other $M-1$ bits are not observed by the single-port detector, each binary hypothesis is a composite hypothesis containing $2^{M-1}$ candidate joint states. We therefore use the profile likelihood over the hidden joint states~\cite{Murphy2000ProfileLikelihood},
\begin{equation}
S_{r,\beta}(y_r;s)
=
\max_{\mathbf{a}\in\mathcal{B}_{r,\beta}}
\mathcal{L}_{s,r}(\mathbf{a};y_r),
\qquad \beta\in\{0,1\}.
\label{eq:single_profile_likelihood}
\end{equation}
The single-port profile-likelihood decision is then
\begin{equation}
\hat b_r(y_r;s)
=
\arg\max_{\beta\in\{0,1\}}S_{r,\beta}(y_r;s).
\label{eq:single_ml}
\end{equation}
The maximization in \eqref{eq:single_profile_likelihood} treats the unobserved $M-1$ bits as nuisance states. Hence, each binary hypothesis is scored by its most likely candidate joint state, rather than by an average over the group. For the $s$th channel realization, the decision region of hypothesis $\beta$ is
\begin{align}
\mathcal{R}_{r,\beta}(s)
&=
\{\,y_r:\,
S_{r,\beta}(y_r;s)\geq S_{r,1-\beta}(y_r;s)
\,\},
\notag\\
&\beta\in\{0,1\}.
\label{eq:single_region}
\end{align}

If the transmitted joint state is $\mathbf{b}$, the conditional bit error probability at port $r$ under the $s$th channel realization is
\begin{equation}
P_{e,r}(\mathbf{b},s)
=
\Pr\left(
Y_r\in\mathcal{R}_{r,1-b_r}(s)
\mid \mathbf{b},s
\right),
\label{eq:single_conditional_ber}
\end{equation}
where $Y_r\mid \mathbf{b},s$ follows the Gaussian model in \eqref{eq:single_conditional_observation}.

Under equal priors over the joint states, the average single-port BER is
\begin{equation}
\mathrm{BER}_r=
\mathbb{E}_{s}
\left[
\frac{1}{|\mathcal{B}|}
\sum_{\mathbf{b}\in\mathcal{B}}
P_{e,r}(\mathbf{b},s)
\right].
\label{eq:single_average_ber}
\end{equation}
In simulations, the expectation over channel realizations and the probability in \eqref{eq:single_conditional_ber} are estimated by Monte Carlo sampling.

\subsection{Joint ML Detection}
The single-port detector uses only one scalar observation. Therefore, the hidden states formed by the other OAM branches may create overlap between the intensity sets associated with $\mathcal{B}_{r,1}$ and $\mathcal{B}_{r,0}$ at port $r$, thereby limiting single-port separability. We therefore consider joint ML detection, which uses the full multi-port observation vector.

For the observation vector $\mathbf{Y}$, the joint detector performs likelihood comparison over the complete joint bit-state set. For an $M$-branch OAM-OOK system, $|\mathcal{B}|=2^M$. Given the same fixed channel realization, the noiseless receiver-side intensity vectors of all candidate joint bit states form the set
\begin{equation}
\mathcal{I}=\{\mathbf{I}(\mathbf{a})\mid \mathbf{a}\in\mathcal{B}\}.
\label{eq:intensity_set}
\end{equation}
If the true transmitted joint bit state is $\mathbf{b}$, then, conditioned on $\mathcal{I}$,
\begin{equation}
\mathbf{Y}\mid \mathbf{b},\mathcal{I}
\sim \mathcal{N}\left(R\mathbf{I}(\mathbf{b}),\boldsymbol{\Sigma}(\mathbf{b})\right).
\label{eq:joint_conditional_observation}
\end{equation}
For any candidate joint bit state $\mathbf{a}$, the conditional density is
\begin{equation}
f_{\mathbf{a}}(\mathbf{Y}\mid \mathbf{I}(\mathbf{a}))
=\mathcal{N}\left(\mathbf{Y};R\mathbf{I}(\mathbf{a}),
\boldsymbol{\Sigma}(\mathbf{a})\right).
\label{eq:joint_pdf}
\end{equation}
Thus, for a fixed conditional intensity set $\mathcal{I}$, the joint ML decision is
\begin{equation}
\hat{\mathbf{b}}(\mathbf{Y};\mathcal{I})
=\arg\max_{\mathbf{a}\in\mathcal{B}}
f_{\mathbf{a}}(\mathbf{Y}\mid \mathbf{I}(\mathbf{a})).
\label{eq:joint_ml}
\end{equation}
Equivalently, after omitting constants independent of the candidate state and multiplying the negative log-likelihood by two, the detector can be written as
\begin{equation}
\hat{\mathbf{b}}(\mathbf{Y};\mathcal{I})
=\arg\min_{\mathbf{a}\in\mathcal{B}}
\Lambda_{\mathbf{a}}(\mathbf{Y};\mathbf{I}(\mathbf{a})),
\label{eq:joint_ml_metric}
\end{equation}
with
\begin{equation}
\Lambda_{\mathbf{a}}(\mathbf{Y};\mathbf{I}(\mathbf{a}))
=\sum_{r=1}^{M}
\frac{(Y_r-RI_r(\mathbf{a}))^2}{\sigma_r^2(I_r(\mathbf{a}))}
+\sum_{r=1}^{M}\log\sigma_r^2(I_r(\mathbf{a})).
\label{eq:joint_metric}
\end{equation}
The likelihood comparison in \eqref{eq:joint_ml} is therefore implemented by the metric minimization in \eqref{eq:joint_ml_metric}, with the heteroscedastic negative log-likelihood metric given by \eqref{eq:joint_metric}. This rule compares the likelihoods of all candidate joint bit states for fixed $\mathcal{I}$.

To define the conditional error probability under fixed $\mathcal{I}$, let the decision region of candidate state $\mathbf{a}$ be
\begin{equation}
\mathcal{R}_{\mathbf{a}}(\mathcal{I})=
\left\{\mathbf{Y}:\Lambda_{\mathbf{a}}(\mathbf{Y};\mathbf{I}(\mathbf{a}))
\leq \Lambda_{\tilde{\mathbf{b}}}(\mathbf{Y};\mathbf{I}(\tilde{\mathbf{b}})),
\ \forall \tilde{\mathbf{b}}\in\mathcal{B}\right\}.
\label{eq:joint_region}
\end{equation}
When the true joint bit state is $\mathbf{b}$, the conditional transition probability of deciding $\mathbf{a}$ is
\begin{equation}
P_{\mathbf{b}\rightarrow\mathbf{a}}(\mathcal{I})
=\int_{\mathcal{R}_{\mathbf{a}}(\mathcal{I})}
f_{\mathbf{b}}(\mathbf{Y}\mid \mathbf{I}(\mathbf{b}))\,\mathrm{d}\mathbf{Y},
\qquad \mathbf{a}\neq\mathbf{b}.
\label{eq:joint_transition}
\end{equation}
The transition probability in \eqref{eq:joint_transition} is computed over the ML decision region defined in \eqref{eq:joint_region}. Under equal priors over joint bit states, the conditional SER is
\begin{equation}
P_{\mathrm{SER}}(\mathcal{I})
=\frac{1}{|\mathcal{B}|}
\sum_{\mathbf{b}\in\mathcal{B}}
\sum_{\substack{\mathbf{a}\in\mathcal{B}\\ \mathbf{a}\neq \mathbf{b}}}
P_{\mathbf{b}\rightarrow\mathbf{a}}(\mathcal{I}).
\label{eq:conditional_ser}
\end{equation}
The corresponding conditional BER weights each erroneous joint decision by the Hamming distance between the transmitted and detected joint states:
\begin{equation}
P_{\mathrm{BER}}(\mathcal{I})
=\frac{1}{M|\mathcal{B}|}
\sum_{\mathbf{b}\in\mathcal{B}}
\sum_{\substack{\mathbf{a}\in\mathcal{B}\\ \mathbf{a}\neq \mathbf{b}}}
d_H(\mathbf{b},\mathbf{a})
P_{\mathbf{b}\rightarrow\mathbf{a}}(\mathcal{I}).
\label{eq:conditional_ber}
\end{equation}
The overall average SER and BER are obtained by averaging the conditional error probabilities over the distribution of the set of receiver-side intensity vectors. Let $p_{\mathcal{I}}(\mathcal{I})$ denote this joint distribution. Then
\begin{equation}
\mathrm{SER}
=\int P_{\mathrm{SER}}(\mathcal{I})p_{\mathcal{I}}(\mathcal{I})\,\mathrm{d}\mathcal{I},
\label{eq:average_joint_ser}
\end{equation}
and
\begin{equation}
\mathrm{BER}
=\int P_{\mathrm{BER}}(\mathcal{I})p_{\mathcal{I}}(\mathcal{I})\,\mathrm{d}\mathcal{I}.
\label{eq:average_joint_ber}
\end{equation}
The conditional error probabilities in \eqref{eq:conditional_ser} and \eqref{eq:conditional_ber} lead to the average SER and BER in \eqref{eq:average_joint_ser} and \eqref{eq:average_joint_ber}, respectively. In simulations, these expectations are approximated by Monte Carlo sampling.

Compared with single-port detection, joint ML detection exploits the complete multi-port observation vector and provides better separability among joint bit states. Therefore, the SCI-D$^2$NN framework developed below is designed to improve the performance of the joint detection.

\section{Supervised Contrastive Inspired D$^2$NN Training Framework}
\label{sec:sci_d2nn_training}

\subsection{Forward Propagation of SCI-D$^2$NN}
We first describe the forward propagation of the SCI-D$^2$NN. As shown in Fig.~\ref{fig:system_framework}, the SCI-D$^2$NN consists of $K$ trainable diffractive planes separated by free-space propagation intervals. Each plane contains diffractive units whose complex transmittance modulates the incident field. The distorted input field $E_{\mathrm{turb}}$ propagates along the $z$ direction and is successively transformed by the transmittance modulation of each plane and the free-space propagation between adjacent planes. Under the scalar diffraction approximation, each diffractive unit can be regarded as a secondary wave source. According to Rayleigh-Sommerfeld diffraction theory, the propagation response observed at spatial position $(x,y,z)$ from the $q$th unit located at $(x_q,y_q,z_q)$ on the $l$th diffractive plane is~\cite{Lin2018AllOpticalD2NN}
\begin{equation}
w_q^l(x,y,z)=
\frac{z-z_q}{r_q^2}
\left(\frac{1}{2\pi r_q}+\frac{1}{j\lambda}\right)
\exp\left(\frac{j2\pi r_q}{\lambda}\right),
\label{eq:rs_node}
\end{equation}
where $r_q=\sqrt{(x-x_q)^2+(y-y_q)^2+(z-z_q)^2}$ and $\lambda$ is the operating wavelength. The complex transmittance of the $q$th unit in the $l$th diffractive plane is $t_q^l=a_q^l\exp(j\phi_q^l)$, where $a_q^l$ and $\phi_q^l$ are the amplitude and phase modulation parameters, typically satisfying $0\leq a_q^l\leq1$ and $0\leq\phi_q^l<2\pi$. The output complex field generated by this unit at a subsequent spatial position can be expressed as
\begin{equation}
U_q^l(x,y,z)=w_q^l(x,y,z)t_q^l
\sum_g U_g^{l-1}(x_q,y_q,z_q),
\label{eq:node_output}
\end{equation}
where the summation term denotes the coherent superposition of the fields incident on this unit from all units in the previous plane.

The unit-level propagation can be equivalently written in a layer-wise complex-field form. The input field of the first layer is $E_{\mathrm{in}}^1(x_1,y_1)=E_{\mathrm{turb}}(x_1,y_1)$. Let $E_{\mathrm{in}}^l(x_l,y_l)$ denote the input field of the $l$th layer. After transmittance modulation, $E^l(x_l,y_l)=E_{\mathrm{in}}^l(x_l,y_l)t^l(x_l,y_l)$. After propagation over the inter-layer distance $d_l$, the output field is
\begin{equation}
E_{\mathrm{out}}^l=
\mathcal{F}^{-1}\left\{
\mathcal{F}\left[E^l(x_l,y_l)\right]H_l(f_x,f_y)
\right\},
\label{eq:fresnel_layer}
\end{equation}
where $\mathcal{F}\{\cdot\}$ and $\mathcal{F}^{-1}\{\cdot\}$ denote the two-dimensional Fourier transform and inverse transform, respectively, and $(f_x,f_y)$ are spatial-frequency coordinates. Under the Fresnel approximation, the propagation function is
\begin{equation}
H_l(f_x,f_y)=
\exp(jkd_l)\exp[-j\pi\lambda d_l(f_x^2+f_y^2)].
\label{eq:fresnel_transfer}
\end{equation}
After sequential transmittance modulation and inter-layer propagation, the SCI-D$^2$NN forms the output field at the output plane. The layer-wise propagation in \eqref{eq:rs_node}--\eqref{eq:fresnel_transfer} realizes the compact SCI-D$^2$NN mapping introduced in \eqref{eq:d2nn_mapping}.

\subsection{Supervised Contrastive Inspired D$^2$NN}

To realize a training objective oriented toward communication-performance optimization, we adopt the principle of supervised contrastive learning, where high-dimensional representations are mapped to a lower-dimensional projection space and label information is used to improve the separability of samples from different classes~\cite{Khosla2020SupCon}.

\begin{figure*}[t]
    \centering
    \includegraphics[width=0.8\textwidth]{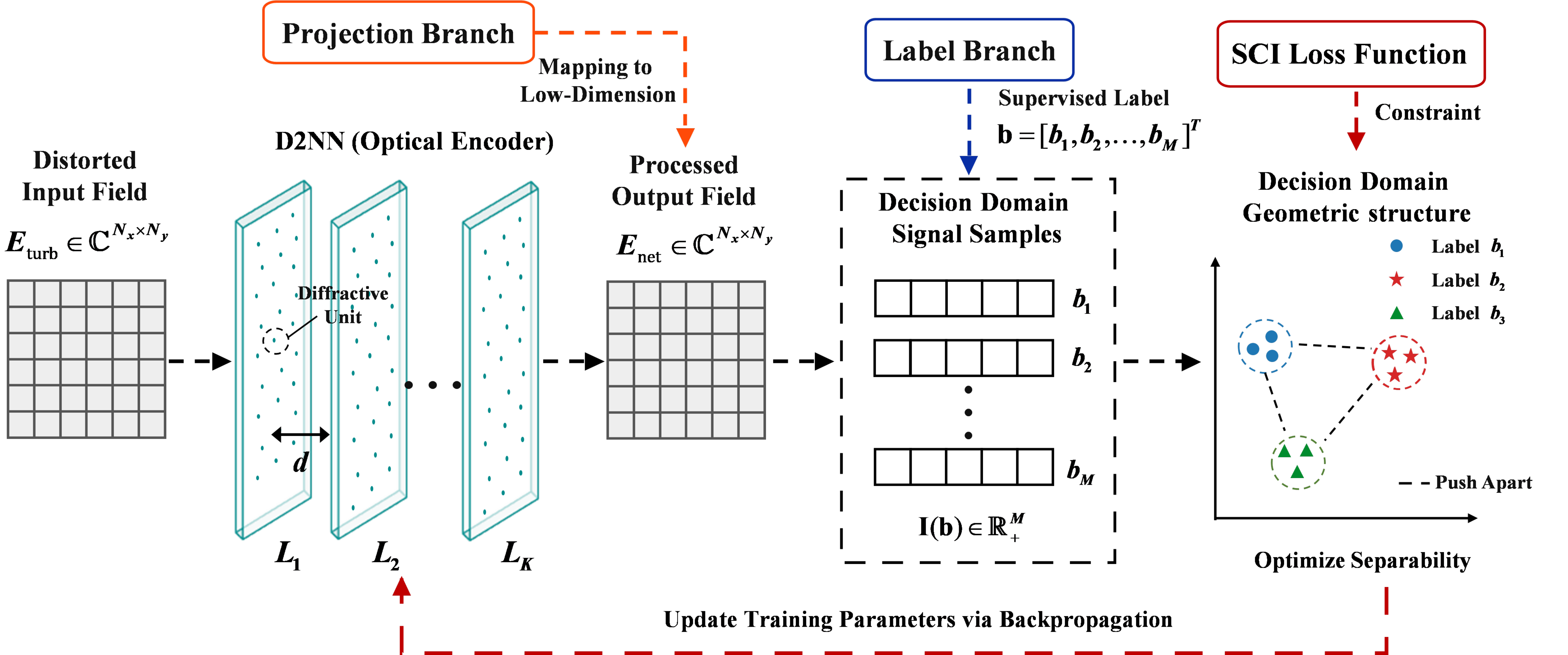}
    \vspace{-0.6em}
    \caption{Training structure of the proposed SCI-D$^2$NN framework}
    \label{fig:sci_d2nn_framework}
\end{figure*}

As illustrated in Fig.~\ref{fig:sci_d2nn_framework}, the proposed SCI-D$^2$NN follows an encoder-projection training structure. The SCI-D$^2$NN optical encoder $F_{\Theta}(\cdot)$ processes the turbulence-distorted complex field $E_{\mathrm{turb}}\in\mathbb{C}^{N_x\times N_y}$ and produces the output field $E_{\mathrm{net}}\in\mathbb{C}^{N_x\times N_y}$, while the receiver-side projection branch maps $E_{\mathrm{net}}$ to the low-dimensional receiver-side intensity vector $\mathbf{I}(\mathbf{b})\in\mathbb{R}_{+}^{M}$ through modal projection, coherent complex-field superposition, and photodetection. The joint bit state $\mathbf{b}$ is used by the label branch as the supervised label for the corresponding decision-domain sample. Since $\mathbf{I}(\mathbf{b})$ determines the receiver observation statistics used by the final ML detector, the SCI-D$^2$NN training constraint is imposed on the geometry of receiver-side intensity vectors in the decision domain to improve their separability and the resulting communication detection performance.

In the proposed SCI-D$^2$NN framework, the joint bit state $\mathbf{b}$ serves as the supervised label. Detection errors tend to occur when the receiver-side intensity vectors of different joint states are close under receiver noise. Among these pairs, Hamming-1 pairs correspond to one-bit error events and are therefore related to BER. We consider two training losses to impose label-guided constraints on the receiver-side decision geometry. Both losses are defined on the receiver-side intensity vectors, but they emphasize different aspects of the joint ML decision structure. The first loss enlarges the Bhattacharyya distance (BD) between Hamming-1 adjacent states, whereas the second uses the joint ML metric to impose separation over all candidate joint states under the short quasi-static channel assumption.

The set of Hamming-1 adjacent joint-state pairs is defined as
\begin{equation}
\mathcal{P}_{\mathrm{adj}}=
\left\{(\mathbf{a},\mathbf{a}'):
\mathbf{a},\mathbf{a}'\in\mathcal{B},\
d_H(\mathbf{a},\mathbf{a}')=1\right\},
\label{eq:adjacent_pairs}
\end{equation}
where $d_H(\cdot,\cdot)$ is the Hamming distance. BD has been widely used as a performance measure in signal selection and in the design of binary or distributed detection systems~\cite{Kailath1967Bhattacharyya,Poor1977AliSilvey,Tarighati2016RateBalancing}. Motivated by this connection, for any adjacent pair $(\mathbf{a},\mathbf{a}')\in\mathcal{P}_{\mathrm{adj}}$, we use BD to measure the receiver-side separability between the two conditional Gaussian observation distributions induced by the two joint states. Let
\begin{equation}
v_r(\mathbf{a})=\sigma_r^2(I_r(\mathbf{a})).
\label{eq:bd_variance}
\end{equation}
The BD is given by
\begin{align}
D_{\mathrm{B}}(\mathbf{a},\mathbf{a}')
&=\sum_{r=1}^{M}\Bigg[
\frac{R^2\left[I_r(\mathbf{a})-I_r(\mathbf{a}')\right]^2}
{4\left[v_r(\mathbf{a})+v_r(\mathbf{a}')\right]}
\notag\\
&\qquad
+\frac{1}{2}
\log\!\left(
\frac{v_r(\mathbf{a})+v_r(\mathbf{a}')}
{2\sqrt{v_r(\mathbf{a})v_r(\mathbf{a}')}}
\right)
\Bigg].
\label{eq:bhattacharyya_distance}
\end{align}
Here, $R$ is the photodetector responsivity, and $v_r(\mathbf{a})$ follows the heteroscedastic photodetection noise model. The first term measures the separation between the conditional means under state-dependent noise, while the second term accounts for the difference between the two noise variances. Therefore, $D_{\mathrm{B}}(\mathbf{a},\mathbf{a}')$ provides a distribution-level measure of the separability between adjacent joint states in the receiver decision space.

For the BD based loss, let $T_{\mathrm{B}}$ be the preset target distance for adjacent joint-state pairs. The value of $T_{\mathrm{B}}$ is specified in the simulation setup. The loss is
\begin{align}
L_{\mathrm{BD}}
&=\mathbb{E}_{\mathrm{batch}}\!\left[
\frac{1}{|\mathcal{P}_{\mathrm{adj}}|}
\sum_{(\mathbf{a},\mathbf{a}')\in\mathcal{P}_{\mathrm{adj}}}
w_{\mathbf{a},\mathbf{a}'}\,
\xi_{\mathbf{a},\mathbf{a}'}
\right],
\label{eq:bd_margin_loss}
\end{align}
where the pairwise margin penalty is defined as
\begin{equation}
\xi_{\mathbf{a},\mathbf{a}'}
=
\operatorname{softplus}\!\left(
\log T_{\mathrm{B}}
-\log D_{\mathrm{B}}(\mathbf{a},\mathbf{a}')
\right).
\label{eq:bd_pairwise_margin_penalty}
\end{equation}
where $|\mathcal{P}_{\mathrm{adj}}|$ is the number of Hamming-1 adjacent pairs, $\mathbb{E}_{\mathrm{batch}}[\cdot]$ denotes empirical averaging over the mini-batch, and $w_{\mathbf{a},\mathbf{a}'}$ is the weight assigned to pair $(\mathbf{a},\mathbf{a}')$. The function $\operatorname{softplus}(x)=\log(1+\exp(x))$ gives a smooth one-sided margin penalty. Therefore, adjacent pairs with $D_{\mathrm{B}}(\mathbf{a},\mathbf{a}')<T_{\mathrm{B}}$ receive larger penalties, whereas pairs that reach the target distance are weakly penalized. The logarithmic form reduces the effect of scale variation among adjacent-pair distances and makes the constraint depend on relative separation.

For the ML based loss, the joint ML negative log-likelihood metric under the short quasi-static channel assumption is used to assign a detection cost to each candidate joint state. For a true joint state $\mathbf{b}$ and a candidate state $\mathbf{a}$, the noiseless receiver statistic $R\mathbf{I}(\mathbf{b})$ is substituted into the candidate-state metric, yielding
\begin{equation}
m_{\mathbf{b},\mathbf{a}}
=
\sum_{r=1}^{M}
\frac{
R^2\left[I_r(\mathbf{b})-I_r(\mathbf{a})\right]^2
}{
\sigma_r^2(I_r(\mathbf{a}))
}
+
\sum_{r=1}^{M}
\log \sigma_r^2(I_r(\mathbf{a})).
\label{eq:ml_training_metric}
\end{equation}
A smaller value of $m_{\mathbf{b},\mathbf{a}}$ indicates that candidate state $\mathbf{a}$ has a lower ML detection cost when the transmitted state is $\mathbf{b}$. We convert this cost into a detection score as
\begin{equation}
s_{\mathbf{b},\mathbf{a}}
=
-\frac{m_{\mathbf{b},\mathbf{a}}}{\tau},
\label{eq:ml_score}
\end{equation}
where $\tau>0$ controls how strongly differences in ML detection cost affect the score. The ML based loss is
\begin{equation}
L_{\mathrm{ML}}
=
\mathbb{E}_{\mathrm{batch}}
\left[
\frac{1}{|\mathcal{B}|}
\sum_{\mathbf{b}\in\mathcal{B}}
-\log
\frac{
\exp(s_{\mathbf{b},\mathbf{b}})
}{
\sum_{\mathbf{a}\in\mathcal{B}}\exp(s_{\mathbf{b},\mathbf{a}})
}
\right].
\label{eq:ml_based_loss}
\end{equation}
The loss penalizes cases in which an incorrect candidate state obtains a detection score comparable to or larger than that of the true state. It therefore imposes ML-metric-based separation over the complete joint-state set.

{
\subsection{Computational Complexity Analysis}

The computational cost of SCI-D$^2$NN training arises from decision-oriented loss evaluation and numerical propagation. For a fixed channel realization in an $M$-branch OAM-multiplexed system employing OOK, the joint-state set contains $|\mathcal{B}|=2^M$ candidate states.

The BD-based loss $L_{\mathrm{BD}}$ evaluates the Bhattacharyya distance over the Hamming-1 adjacent-pair set $\mathcal{P}_{\mathrm{adj}}$, which contains $M2^{M-1}$ pairs. Since each distance evaluation involves $M$ receiver ports, its core computational complexity is $O(M^2 2^{M-1})$ for each channel realization. In contrast, the ML-based loss $L_{\mathrm{ML}}$ evaluates the receiver-side ML metric for every transmitted-state and candidate-state combination. Its computational complexity is therefore $O(M|\mathcal{B}|^2)=O(M4^M)$ for each channel realization. Thus, $L_{\mathrm{BD}}$ has a more favorable scaling with the number of multiplexed OAM branches.

During forward and backward propagation, the main numerical cost arises from FFT-based angular-spectrum propagation. For a single complex-field input, each free-space propagation interval requires a two-dimensional FFT and inverse FFT, with computational complexity $O(N_xN_y\log(N_xN_y))$ for an $N_x\times N_y$ complex-field grid. Constructing the crosstalk matrix for one channel realization requires calculating the response fields of the $M$ unit-amplitude OAM branches. Since each field propagates through $K+1$ intervals, the corresponding forward-propagation complexity is $O\!\left(M(K+1)N_xN_y\log(N_xN_y)\right)$. The corresponding backward propagation has a comparable computational cost.

For the considered three-branch configuration, FFT-based angular-spectrum propagation constitutes the main training cost because of the large complex-field grid. As $M$ increases, the computational complexities of both loss functions grow exponentially, whereas the numerical-propagation complexity increases linearly with $M$. For a fixed field-grid size and network depth, the loss evaluation can eventually exceed the numerical-propagation cost and become the dominant computational cost during training.

\subsection{Scalability Discussion}

The above complexity analysis motivates further consideration of the proposed framework for larger OAM-multiplexed systems. Practical high-capacity FSO applications further require the framework to support higher-order modulation formats and adapt to stronger inter-mode crosstalk and more complex channel conditions. The feasibility of hardware deployment also needs to be considered.

For larger OAM-multiplexed systems, the main SCI-D$^2$NN architecture can be retained, but low-complexity training strategies are needed to reduce the loss-evaluation cost. A groupwise strategy divides the OAM branches into smaller groups and optimizes the joint states within each group, thereby reducing the joint-state dimension considered in each training step~\cite{Qiu2023GroupwiseDetection}. In addition, a list-based strategy retains the complete OAM multiplexing dimension but restricts loss evaluation to a finite list of representative or competing joint states~\cite{Nammi2009ListBasedDetection}. Both approaches can reduce the exponential growth in the computational complexity of the two loss functions for larger OAM-multiplexed systems, although they introduce a performance-complexity trade-off.

For higher-order modulation formats requiring coherent reception, such as QPSK and QAM, the intensity-based direct-detection receiver and training model of the proposed framework need to be extended to a complex decision domain. For a given channel realization $s$, the coherent receiver observation is
\begin{equation}
\mathbf{z}_s=\mathbf{A}_s\mathbf{P}^{1/2}\mathbf{c}+\mathbf{n},
\end{equation}
where $\mathbf{z}_s$ is the complex observation vector and $\mathbf{c}$ denotes the transmitted joint QPSK/QAM constellation state. The noise at receiver port $r$ is expressed as $n_r=n_{I,r}+j n_{Q,r}$, where $n_{I,r}$ and $n_{Q,r}$ are the in-phase and quadrature noise components, respectively, and satisfy $n_{I,r},n_{Q,r}\sim\mathcal{N}(0,\sigma_c^2)$. Under the common strong-local-oscillator coherent-receiver model~\cite{Bayaki2012CoherentDifferential},
\begin{align}
\sigma_c^2 &= \sigma_{\mathrm{sh}}^2+\sigma_{\mathrm{th}}^2,\notag\\
\sigma_{\mathrm{sh}}^2 &= 2q_eRP_{\mathrm{LO}}B_e,\notag\\
\sigma_{\mathrm{th}}^2 &= \frac{4k_{\mathrm{B}}TB_e}{R_L},
\end{align}
where $\sigma_c^2$ is the total noise variance of each quadrature component and $P_{\mathrm{LO}}$ is the local-oscillator power. When $P_{\mathrm{LO}}$ is much larger than the received signal power $P_{\mathrm{s}}$, shot noise is dominated by the local oscillator and is approximately independent of the transmitted constellation state. The noise vector can then be modeled as $\mathbf{n}\sim\mathcal{CN}(\mathbf{0},2\sigma_c^2\mathbf{I}_M)$, where $\mathbf{I}_M$ is the $M\times M$ identity matrix.

Accordingly, the joint ML detector is extended from the OOK joint-state set to the joint QPSK/QAM constellation set. For a candidate joint state $\mathbf{a}$, its noiseless complex observation is $\boldsymbol{\mu}_{s,\mathbf{a}}=\mathbf{A}_s\mathbf{P}^{1/2}\mathbf{a}$, and the ML metric becomes
\begin{equation}
\Lambda_{\mathbf{a}}(\mathbf{z}_s)
=
\frac{\left\lVert\mathbf{z}_s-\boldsymbol{\mu}_{s,\mathbf{a}}\right\rVert_2^2}
{2\sigma_c^2}.
\end{equation}
The softmax structure of $L_{\mathrm{ML}}$ can be retained by replacing the intensity-domain detection cost with $m_{\mathbf{c},\mathbf{a}}=\left\lVert\boldsymbol{\mu}_{s,\mathbf{c}}-\boldsymbol{\mu}_{s,\mathbf{a}}\right\rVert_2^2/(2\sigma_c^2)$, while preserving the same detection-score and softmax-loss construction.

For $L_{\mathrm{BD}}$, the equal-covariance complex Gaussian model gives the Bhattacharyya distance between two joint constellation states as
\begin{equation}
D_{\mathrm{B}}(\mathbf{a},\mathbf{a}')
=
\frac{\left\lVert\boldsymbol{\mu}_{s,\mathbf{a}}
-\boldsymbol{\mu}_{s,\mathbf{a}'}\right\rVert_2^2}
{8\sigma_c^2}.
\end{equation}
The OOK Hamming-1 adjacent-pair set $\mathcal{P}_{\mathrm{adj}}$ is replaced by joint QPSK/QAM constellation pairs in which one OAM branch undergoes a nearest-neighbor constellation transition under Gray labeling. These modifications enable the proposed framework to support higher-order modulation formats. The number of joint candidate states then increases from $2^M$ for OOK to $4^M$ for QPSK and $q^M$ for $q$-QAM, leading to higher training cost.

For practical FSO communication systems, the proposed framework needs to address stronger inter-mode crosstalk and more complex channel conditions. To improve its adaptability to these conditions, more accurate FSO channel models can be used to generate training samples. For example, existing models account for platform jitter and tracking noise~\cite{Shang2025OISLChannel}, as well as complex effects such as particle scattering, turbulence, and multi-medium propagation~\cite{Wang2026SatelliteUnderwater}. By more completely characterizing wavefront variations in the received optical field, these models can reduce the mismatch between the propagation model used during training and an actual FSO link. Besides, expected assembly and fabrication errors can be incorporated as random perturbations in the optical forward model during robust training~\cite{Mengu2020Misalignment}, thereby improving the tolerance of SCI-D$^2$NN to practical device imperfections. In addition to these training-stage measures, receiver-side signal-processing schemes, e.g., equalization, may be combined with SCI-D$^2$NN to further improve its applicability in practical FSO communication systems.

When deploying the proposed SCI-D$^2$NN framework, the projection and label branches are used only during offline training, and physical implementation therefore requires only the trained SCI-D$^2$NN optical front end. The diffractive layers can be implemented using spatial light modulators or diffractive optical elements (DOEs), e.g., metasurfaces. These diffractive layers jointly realize optical inference through cascaded diffraction propagation. To reduce discrepancies between the physical system and the numerical model, the implemented diffractive optical elements should be accurately calibrated to match the geometric configuration used during training. Existing multilayer diffractive optical experiments have adopted CCD-based real-time imaging for layer-by-layer calibration~\cite{Fan2024HolographicMultiplexing}. In this procedure, the reference spot is first centered using CCD imaging, and subsequent diffractive layers are aligned layer by layer using the previous layer as the reference, thereby correcting relative inter-layer position errors.
}

\subsection{Traditional D$^2$NN for Comparison}

To benchmark the proposed decision-oriented training, we include the conventional field-restoration D$^2$NN reported in~\cite{Jia2022D2NNDelay}. The network uses the optical front end in \eqref{eq:d2nn_mapping}, but the supervision is imposed on the output field itself. It takes a turbulence-free reference field as the label; consequently, the loss does not directly act on the projected intensity vectors or the ML decision metric.
For the $n$th training sample, let $E_{\mathrm{net},n}$ and $E_{\mathrm{ideal},n}$ denote the D$^2$NN output field and the corresponding ideal target field, respectively. The traditional complex-field restoration loss is
\begin{align}
L_{\mathrm{baseline}}
&=\frac{1}{N}\sum_{n=1}^{N}
\left(E_{\mathrm{ideal},n}
-\frac{e^{j\varphi_n}E_{\mathrm{net},n}}{t_0}\right)^{\dagger}
\notag\\
&\quad \times
\left(E_{\mathrm{ideal},n}
-\frac{e^{j\varphi_n}E_{\mathrm{net},n}}{t_0}\right),
\label{eq:jia_loss}
\end{align}
Here, $t_0$ is the preset target transmittance. The phase factor is defined as
\begin{equation}
e^{j\varphi_n}
=\frac{
E_{\mathrm{net},n}^{\dagger}E_{\mathrm{ideal},n}
}{
\left|E_{\mathrm{net},n}^{\dagger}E_{\mathrm{ideal},n}\right|
}
\label{eq:jia_phase}
\end{equation}
The scalar phase factor in \eqref{eq:jia_phase} removes the sample-dependent global phase ambiguity between the output and reference fields. With this alignment and the target transmittance $t_0$, \eqref{eq:jia_loss} penalizes the residual complex-field mismatch. We use this objective as the field-restoration baseline.

{
Compared with the proposed SCI-D$^2$NN, the traditional D$^2$NN baseline in the following simulations includes a diaphragm. Consistent with the field-restoration configuration of Jia \emph{et al.}, the diaphragm suppresses stray light and thereby improves the effectiveness of field-restoration training. We found through numerical simulations that the traditional D$^2$NN trained without the diaphragm yields higher BER under both joint ML and single-port PL detection. We therefore use the diaphragm-equipped traditional D$^2$NN as the comparison baseline.
}

\section{Numerical Results}
\label{sec:numerical_results}

In this section, we evaluate the improvement achieved by the proposed SCI-D$^2$NN in the detection performance of OAM-multiplexed FSO communications under atmospheric turbulence, transmitter pointing errors, and photodetection noise. Unless otherwise specified, the main simulation parameters are summarized in Table~\ref{tab:simulation_parameters}. The OAM mode set is $S=\{1,3,5\}$, corresponding to $|\mathcal{B}|=2^M=8$ joint OOK states. Three turbulence strengths, \(C_n^2=5\times10^{-14}\), \(10^{-13}\), and \(2\times10^{-13}~\mathrm{m}^{-2/3}\), are considered in the simulations. {The SCI-D$^2$NN models used in these simulations were implemented in Python 3.12.8 and trained using PyTorch (version 2.7.0 with Compute Unified Device Architecture (CUDA) 11.8) on a single NVIDIA GeForce RTX 4090 graphics processing unit (GPU). The five-layer SCI-D$^2$NN has 800,000 trainable phase parameters. Each training epoch consisted of 300 iterations of gradient descent and took approximately 28.0 s using the $L_{\mathrm{BD}}$ objective and 30 s using $L_{\mathrm{ML}}$. The passive inference delay of the diffractive optical front end is $t_{\mathrm{prop}}=(K+1)d/c$, which is approximately $1.0~\mathrm{ns}$ for $K=5$ and $d=5~\mathrm{cm}$.}

\begin{table}[!t]
\caption{Simulation Parameters}
\label{tab:simulation_parameters}
\centering
\begin{tabular}{p{0.24\textwidth}p{0.20\textwidth}}
\hline
\hline
\textbf{Parameter} & \textbf{Value} \\
\hline
OAM mode set, \(S\) & \(\{1,3,5\}\) \\
Wavelength, \(\lambda\) & \(1550~\mathrm{nm}\) \\
Propagation distance, \(d\) & \(1000~\mathrm{m}\) \\
Number of phase screens, \(N_p\) & \(10\) \\
Outer scale, \(L_0\) & \(10~\mathrm{m}\) \\
Inner scale, \(l_0\) & \(0.01~\mathrm{m}\) \\
Complex-field samples, \(N_x\times N_y\) & \(400\times400\) \\
Default symbol rate, \(R_s\) & \(1~\mathrm{Gbaud}\) \\
Receiver noise bandwidth, \(B_e\) & \(1~\mathrm{GHz}\) \\
Pointing-error standard deviation per axis, \(\sigma_\theta\) & \(100~\mu\mathrm{rad}\) \\
Pointing offset variance & \(1.0\times10^{-2}~\mathrm{m^2}\) per axis \\
Number of diffractive layers, \(K\) & \(5\) \\
Distance between layers & \(5~\mathrm{cm}\) \\
Trainable SCI-D$^2$NN parameters & Phase-only masks \\
Optimizer & AdamW \\
\hline
\hline
\end{tabular}
\end{table}

The raw distorted field and the SCI-D$^2$NN input field are sampled on different grids. The raw distorted field uses $\Delta x_{\mathrm{raw}}=1~\mathrm{mm}$, whereas the SCI-D$^2$NN field uses $\Delta x_{\mathrm{D^2NN}}=10~\mu\mathrm{m}$. Before entering the SCI-D$^2$NN, the field amplitude is scaled by a factor of $100$ to keep the optical-power normalization consistent across the two sampling grids. This scaling corresponds to the lens-assisted focusing assumed between the propagation plane and the SCI-D$^2$NN input plane. With equal prior probabilities for the OOK symbols, the on-state power is $P_{\mathrm{on}}=P_{\mathrm{avg}}/1.5$ for $M=3$. The receiver noise bandwidth \(B_e\) is set to the symbol-rate.

\begin{figure}[!tbp]
\centering
\includegraphics[width=\columnwidth]{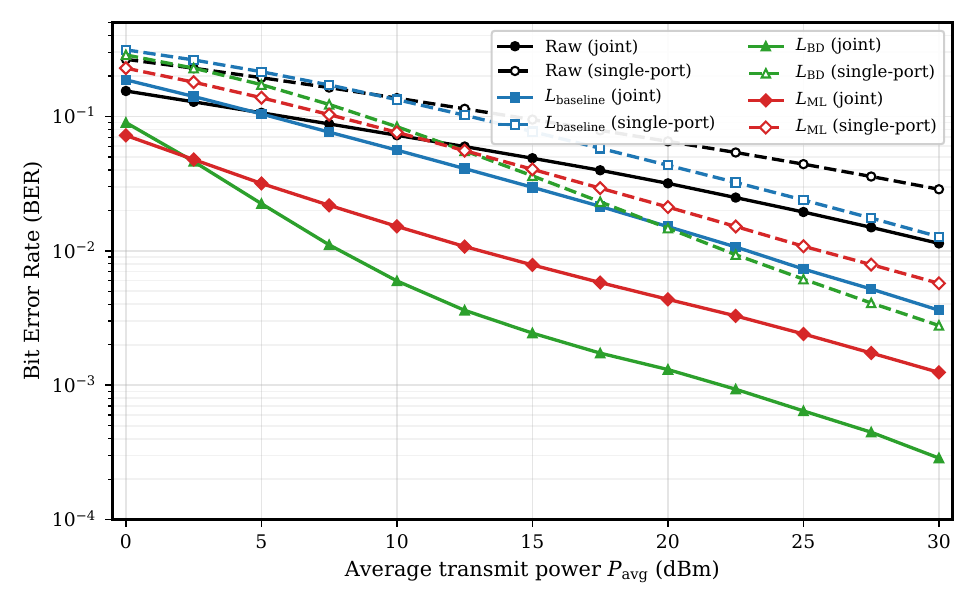}
\caption{BER comparison between single-port profile-likelihood detection and joint ML detection.}
\label{fig:sigma100_joint_vs_single_port_ml_ber}
\end{figure}

We first present the BER comparison between single-port PL detection and joint ML detection in Fig.~\ref{fig:sigma100_joint_vs_single_port_ml_ber}. First, we observe that the three-port joint ML detector consistently outperforms the corresponding single-port detector, since it exploits the multidimensional receiver observations rather than relying on one port only. The following BER simulations therefore use the three-port joint ML detector as the main detection protocol. Besides, both SCI-D$^2$NN training schemes provide clear BER advantages over the raw received field and the conventional D$^2$NN baseline, with more than a 3-dB transmit-power gain over the baseline in most transmit-power regions. {This improvement may come from the receiver-side projection branch and decision-oriented training objective, which map the optical field to low-dimensional decision-domain samples and optimize their separability according to the final detection criterion.}

\FloatBarrier

\begin{figure}[!tbp]
\centering
\includegraphics[width=\columnwidth]{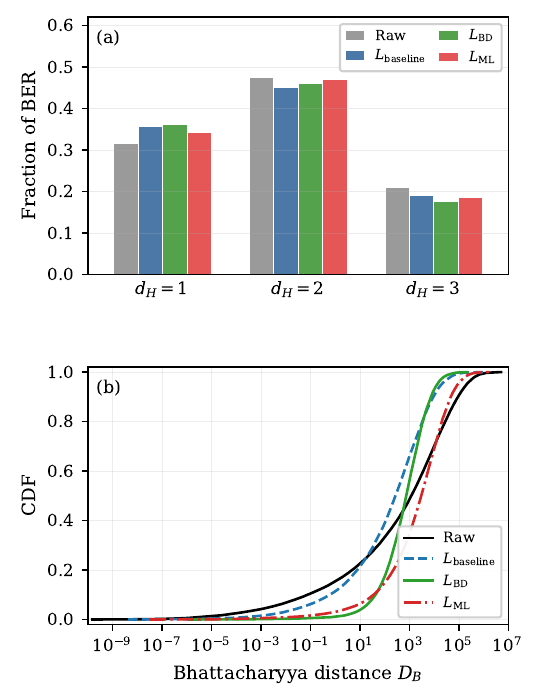}
\caption{BER-error composition and decision-domain separation at \(P_{\mathrm{avg}}=10~\mathrm{dBm}\): (a) fraction of BER contributed by pairs with different Hamming distances \(d_H=1,2,3\); (b) cumulative distribution functions (CDFs) of the BD values for joint-state pairs.}
\label{fig:error_fraction_and_bd_cdf}
\end{figure}

Then, we present the BER contribution ratios of different Hamming-distance transitions in Fig.~\ref{fig:error_fraction_and_bd_cdf}(a). First, we observe that the BER contribution is not dominated only by nearest Hamming-1 errors; for all four conditions, \(d_H=2\) transitions account for the largest BER fraction. This indicates that higher Hamming distance error events cannot be ignored when evaluating joint detection performance. From Fig.~\ref{fig:sigma100_joint_vs_single_port_ml_ber} and Fig.~\ref{fig:error_fraction_and_bd_cdf}(a), it is worth noting that \(L_{\mathrm{BD}}\), although explicitly imposed only on Hamming-1 adjacent state pairs, still improves the separability of larger-\(d_H\) state pairs. Moreover, the Hamming-1 adjacent-pair formulation provides a lower-complexity training surrogate while retaining the main BER improvement. Furthermore, the BER contribution of \(d_H=3\) transitions is relatively small, which may be attributed to the fact that they require three simultaneous bit errors and are therefore less likely to occur under the same channel realization. To further explain the BER behavior in Fig.~\ref{fig:sigma100_joint_vs_single_port_ml_ber}, we present the distribution of the Bhattacharyya distance \(D_{\mathrm{B}}\) between joint-state observation distributions in Fig.~\ref{fig:error_fraction_and_bd_cdf}(b). First, we observe that the raw distorted field has a pronounced low-\(D_{\mathrm{B}}\) tail. This explains its high BER, since these overlapping conditional distributions are difficult to distinguish by the joint ML detector. Besides, the conventional D$^2$NN baseline only moderately improves the weak-separation tail and does not consistently shift the main distribution toward larger \(D_{\mathrm{B}}\), which explains its limited BER gain. Moreover, the two SCI-D$^2$NN losses improve the decision-domain separability more effectively: \(L_{\mathrm{BD}}\) provides the strongest correction of weakly separated state pairs and therefore achieves the lowest BER over most of the considered power range, whereas \(L_{\mathrm{ML}}\) imposes a broader multi-class constraint over all joint states and can be more balanced in the low power region. The difference between the two losses is therefore consistent with their objectives, with \(L_{\mathrm{BD}}\) focusing on the weak-separation tail and \(L_{\mathrm{ML}}\) improving the overall multi-class likelihood geometry.

\FloatBarrier

\begin{figure}[!tbp]
\centering
\includegraphics[width=\columnwidth]{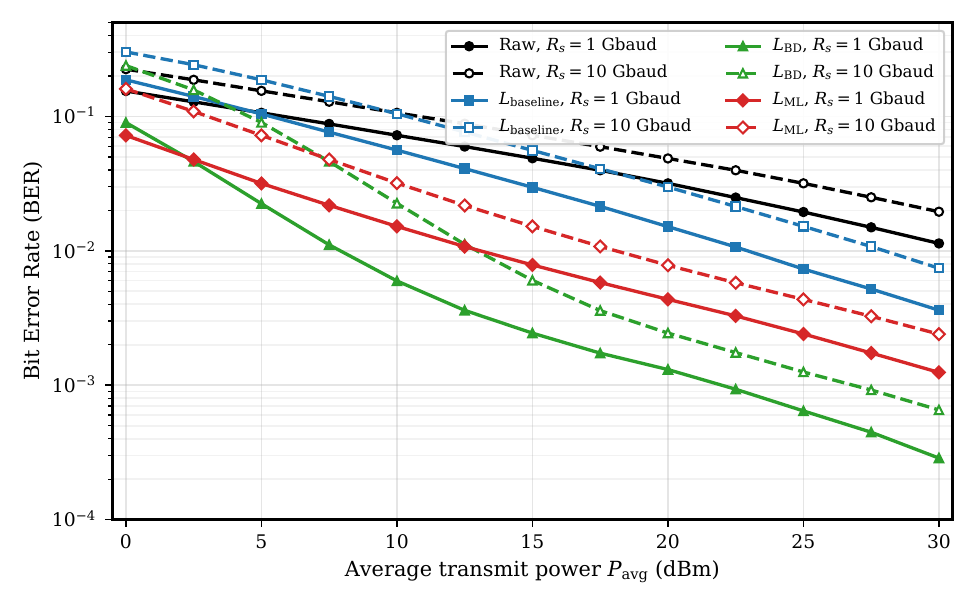}
\caption{BER comparison for \(R_s=1\) and \(10\) Gbaud.}
\label{fig:bandwidth_sweep_joint_ml}
\end{figure}

Then, we explore the BER performance under different symbol rates in Fig.~\ref{fig:bandwidth_sweep_joint_ml}. First, we observe that both SCI-D$^2$NN losses still provide clear BER improvements over the raw distorted field and the conventional D$^2$NN baseline when the symbol rate increases, indicating that the detection-oriented training remains effective under a larger receiver bandwidth and increased noise variance. Besides, as the symbol rate increases, the transmit-power region where \(L_{\mathrm{ML}}\) outperforms \(L_{\mathrm{BD}}\) becomes wider, which is consistent with the observation that \(L_{\mathrm{ML}}\) imposes a more balanced multi-class constraint on the decision geometry. Moreover, \(L_{\mathrm{BD}}\) still achieves better detection performance in most transmit-power regions, because its lower-tail correction remains effective for detection performance. Therefore, \(L_{\mathrm{BD}}\) is used as the representative SCI-D$^2$NN loss in the following simulations.

\FloatBarrier

\begin{figure}[!tbp]
\centering
{
\subfigure[]{
\includegraphics[width=0.98\columnwidth]{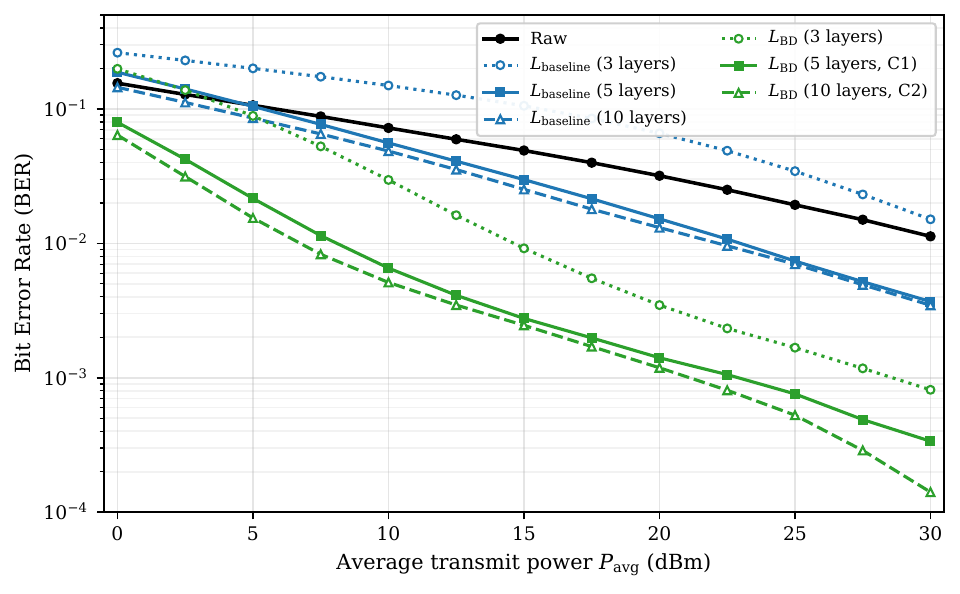}
}\\[-0.5em]
\subfigure[]{
\includegraphics[width=0.98\columnwidth]{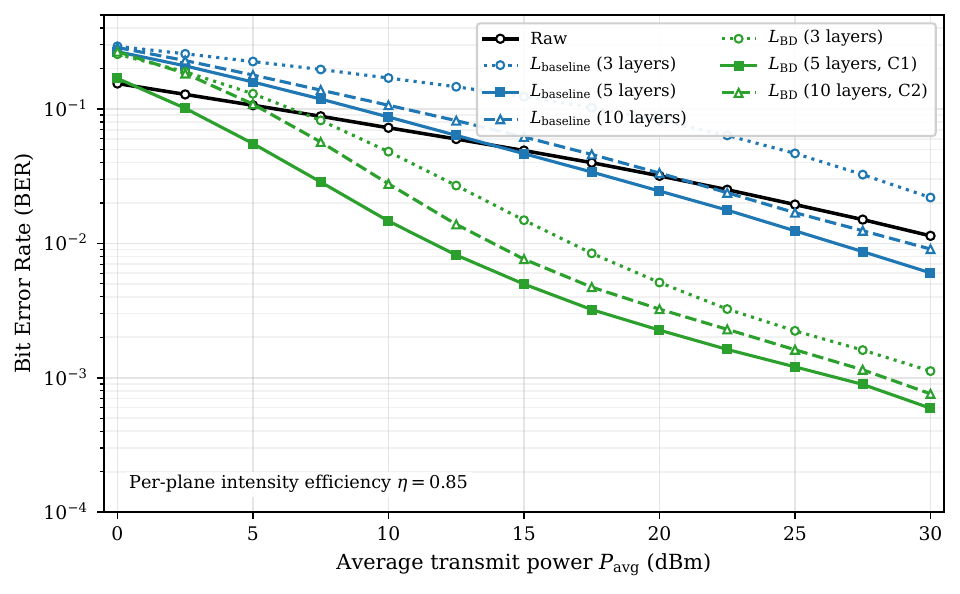}
}
\caption{BER comparisons under different diffractive layers: (a) ideal condition without diffraction loss; (b) per-layer intensity diffraction efficiency of \(\eta=0.85\).}
\label{fig:layer_ablation_joint_ml}
}
\end{figure}

{
Then, we present the BER comparisons under different diffractive layers in Fig.~\ref{fig:layer_ablation_joint_ml}. Under the ideal condition without diffraction loss in Fig.~\ref{fig:layer_ablation_joint_ml}(a), both the traditional D$^2$NN baseline and SCI-D$^2$NN improve as the number of diffractive layers increases, with a clear gain from three to five layers but only a limited additional gain from five to ten layers. This indicates that additional diffractive layers provide more optical degrees of freedom for training, while the marginal benefit gradually decreases. Besides, $L_{\mathrm{BD}}$ shows a much larger performance advantage: even with three layers, it already achieves good detection performance, and in the high-power region it outperforms the ten-layer baseline. This suggests that SCI-D$^2$NN can better exploit the capability of the network by using the projection branch to simplify the training task and by applying a loss function aligned with detection performance. However, Fig.~\ref{fig:layer_ablation_joint_ml}(a) assumes lossless diffraction. In practical implementations, finite diffraction efficiency leads to cumulative optical loss across multiple diffractive layers. Therefore, Fig.~\ref{fig:layer_ablation_joint_ml}(b) considers a loss-aware setting with a typical per-layer intensity diffraction efficiency of \(\eta=0.85\)~\cite{Yang2019LCOSOpticalEfficiency}, where the complex field is multiplied by \(\sqrt{\eta}\) after each diffractive modulation. Under this loss-aware condition, the five-layer SCI-D$^2$NN achieves the lowest BER among the considered depth settings. Although the three-layer design preserves more optical power, its optical processing capability is less effective, whereas the limited gain of the ten-layer design without diffraction loss is offset by accumulated diffraction loss. Therefore, the five-layer SCI-D$^2$NN is adopted as the default configuration in the subsequent simulations.
}

\FloatBarrier

\begin{figure}[!tbp]
\centering
\includegraphics[width=\columnwidth]{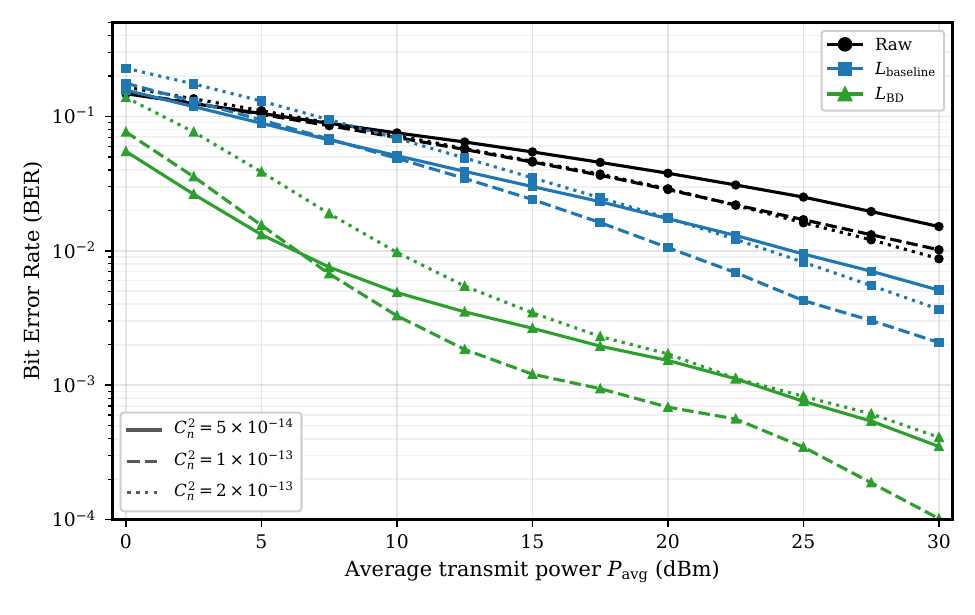}
\caption{BER comparison under different turbulence strengths.}
\label{fig:sigma100_per_turbulence_joint_ml}
\end{figure}

Then, we explore the effect of turbulence strength on the BER performance. Fig.~\ref{fig:sigma100_per_turbulence_joint_ml} compares the BER curves under three different turbulence strengths. As the turbulence strength increases, the raw distorted field gives higher BER. Under different turbulence strengths, \(L_{\mathrm{BD}}\) consistently improves the detection performance compared with the raw distorted field and \(L_{\mathrm{baseline}}\). It is worth noting that the case of \(C_n^2=1\times10^{-13}\) gives lower BER over most transmit-power points. This non-monotonic behavior may be related to the mixed-turbulence training strategy. During mixed training, the D$^2$NN learns a shared solution across the three turbulence conditions, and this solution may be better matched to the \(C_n^2=1\times10^{-13}\) case, resulting in a more favorable decision-domain geometry.

\FloatBarrier

\begin{figure}[!tbp]
\centering
\includegraphics[width=\columnwidth]{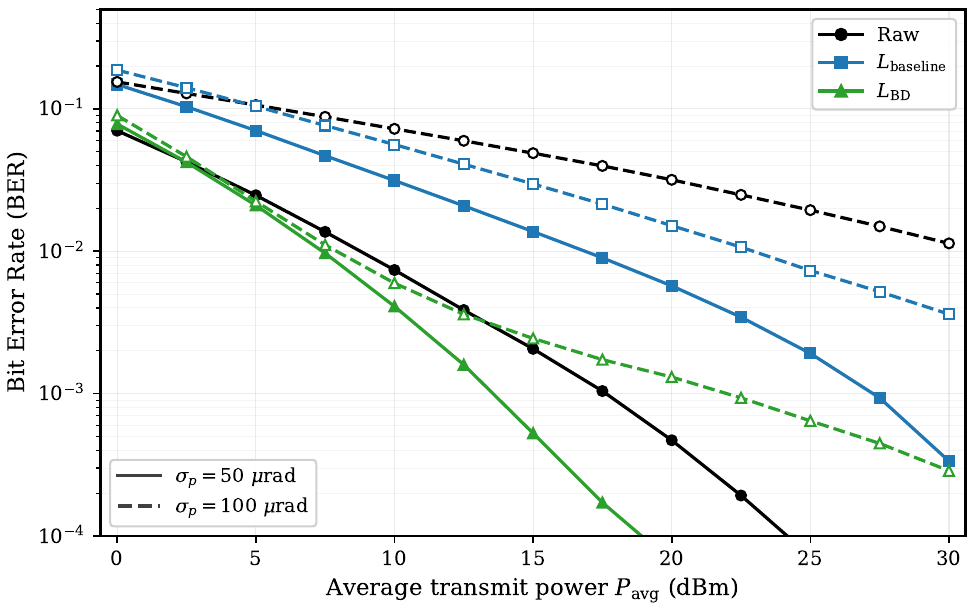}
\caption{BER comparison under different pointing-errors.}
\label{fig:pointing_error_comparison_lbd}
\end{figure}

Finally, we present the BER comparison under different pointing-error levels in Fig.~\ref{fig:pointing_error_comparison_lbd}. First, we observe that when the pointing error is reduced, the conventional D$^2$NN baseline fails to improve the detection performance and can even give worse BER than the raw distorted field. However, \(L_{\mathrm{BD}}\) still provides clear detection performance gain over most transmit power regions, showing that the BD based SCI-D$^2$NN loss remains effective under different pointing-error conditions. Moreover, together with the previous BER results, this comparison shows that \(L_{\mathrm{BD}}\) provides robust detection performance improvement under different system parameters, achieving more than a 10-dB BER gain over the conventional D$^2$NN baseline in the high transmit power region.

\FloatBarrier
\clearpage

\section{Conclusion}
\label{sec:conclusion}
Existing D$^2$NN-based compensation schemes for turbulence-distorted OAM beams did not consider detection-performance optimization for communication. To address this limitation, we proposed an SCI-D$^2$NN framework with two training branches for improving the detection performance of OAM-multiplexed FSO communication under atmospheric turbulence, transmitter pointing errors, and photodetection noise. The projection branch simplifies the training task by mapping the SCI-D$^2$NN output field to low-dimensional receiver-side intensity samples. The label branch introduces the transmitted joint symbol states as supervised labels to improve the decision-domain separability of samples according to the symbol-state relations. In addition, receiver-side complex-amplitude crosstalk was characterized to obtain the observation vector used for detection. Then, single-port profile-likelihood and joint ML detection schemes were formulated, and the corresponding BER and SER criteria were derived. Moreover, we designed two loss functions for SCI-D$^2$NN to optimize the detection performance. Numerical results demonstrated that the proposed SCI-D$^2$NN framework achieves more than a 3-dB BER improvement over the conventional D$^2$NN baseline in most transmit-power regions by reducing the dimensionality of the training samples and aligning the optimization objective with the final detection performance. Among the proposed losses, the BD based loss achieves the lowest BER under different system parameters, including symbol rate, turbulence strength, number of diffractive layers, and pointing error, and provides more than a 10-dB BER improvement over the conventional D$^2$NN baseline in the high-transmit-power region by effectively improving the separability of weakly separated samples. Our work sheds light on the use of D$^2$NNs to improve communication-system performance.

\bibliographystyle{IEEEtran}
\bibliography{SCI-D2NN-revised-v2}

\end{document}